\documentclass[journal]{IEEEtran}
\IEEEoverridecommandlockouts
\usepackage{cite}
\usepackage{amsmath,amssymb,amsfonts}
\usepackage{algorithmic}
\usepackage{graphicx}
\usepackage{textcomp}
\usepackage{xcolor}
\usepackage{todonotes}
\usepackage{subcaption}

\usepackage{background}

\newcommand{\ffig}[1]{Figure~\ref{#1}}

\backgroundsetup{
  contents=IEEE SAS 2024 submission: \#1571011682,
  color=black,
  opacity=1,
  angle=0,
  scale=1,
  position=current page.north east,
  hshift=-5cm,
  vshift=-1cm}

\def\BibTeX{{\rm B\kern-.05em{\sc i\kern-.025em b}\kern-.08em
    T\kern-.1667em\lower.7ex\hbox{E}\kern-.125emX}}
\begin{document}

\title{Real-time smart self-mixing interferometry sensor with embedded neural network

\thanks{This project has received financial support from the CNRS through the MITI interdisciplinary programs, the ANR PEPR IA Emergences project ANR-23-PEIA-0002 and  by the French government, through the France 2030 investment plan managed by the Agence Nationale de la Recherche, as part of the Université Côte d'Azur's Initiative of Excellence, reference ANR-15-IDEX-0001.}
}

\author{Pierre-Emmanuel Novac, Laurent Rodriguez, and Stéphane Barland
\thanks{P.-E. Novac and L. Rodriguez are with Université Côte d'Azur, LEAT, Sophia Antipolis, France (e-mail: pierre-emmanuel.novac@univ-cotedazur.fr; laurent.rodriguez@univ-cotedazur.fr)}
\thanks{S. Barland is with INPHYNI, CNRS, Université Côte d'Azur, Nice, France (e-mail: stephane.barland@univ-cotedazur.fr)}
}

\maketitle

\begin{abstract}
  Self-mixing interferometry is a measurement approach in which a laser beam is re-injected into the emitting laser itself after reflection on a target.
  Information about the position of the target can be obtained from monitoring the voltage across the laser.
  However, analyzing this signal is difficult.
  In previous works, neural networks have been used with great success to process this data.
  In this article, we present an updated prototype of an integrated sensor based on self-mixing interferometry with embedded neural networks.
  It consists of a semiconductor laser (acting both as light emitter and detector) equipped with an embedded platform for data processing.
  The platform includes an ADC (Analog-to-Digital Converter) and an STM32U575ZI microcontroller.
  The microcontroller runs a residual neural network in charge of reconstructing the displacement of a target from the interferometric signal entering the ADC.
  We assess the robustness of the neural network to unwanted signal amplitude variations,
  the impact of different network weights quantization choices required to run the network on the microcontroller,
  and the energy consumption of the system.
  Finally, we provide a demonstration of target displacement reconstruction fully running on the embedded platform in real-time.
  Our results pave the way towards robust, low power and versatile sensors based on self-mixing interferometry and embedded neural networks.

\end{abstract}

\begin{IEEEkeywords}
Self-mixing interferometry, semiconductor lasers, neural networks, embedded artificial intelligence 
\end{IEEEkeywords}

\section{Introduction}

Self-mixing interferometry is a well-established and versatile measurement approach in which a laser beam is re-injected into the emitting laser itself after reflection on a target \cite{giuliani2002laser,kane2005unlocking,donati2012developing,taimre2016laser,li2017laser,rakic2019sensing}.
Thus, optical interferences between the emitted and reflected beams take place inside the laser itself, altering its operation point.
In normal operating conditions, a semiconductor laser emits a monochromatic beam whose intensity (and all other properties) are fully determined by the laser construction parameters and by the amount of energy brought into the laser, \textit{ie} the electrical pumping current.
In the case of self-mixing interferometry, the interference condition between the emitted and the re-injected beam constitutes an additional parameter which influences the intensity emitted by the laser.
Thus, information about the position of the target (or many other properties, see \textit{eg} \cite{brambilla2020versatile} for an overview) can be obtained from monitoring the laser operation point, by recording the voltage across the laser diode as a function of time. The physics of laser self-mixing is well described by the Lang and Kobayashi equations \cite{lang1980external}, in which the reflection of the beam on a target of any material is simply embedded in a delayed feedback term whose amplitude is set by the reflectivity of the target material \cite{giuliani2002laser}. The operating range of the system is well described by the so-called feedback parameter $C \propto \frac{\kappa s \sqrt{1+\alpha^2}}{Ln}$ where $\alpha,L,n$ are physical parameters of the laser (amplitude-phase coupling, length and refractive index respectively) and $\kappa,s$ describe the effective target reflectivity and distance. Precise and exhaustive information about operating parameters can be found for instance in \cite{giuliani2002laser} and in practical terms the approach works well for values of $C$ near unity, which brings target reflectivities and distance of the order of $10^{-4}$ and few tens of cm respectively.
Recovering the displacement of the target per time unit is in principle a simple inverse problem which can be solved with two different approaches depending on the displacement range.
If the displacement is less than an optical wavelength, the analytic expression of the stationary states of the Lang and Kobayashi model can be fit to the data and the displacement can be recovered (provided other parameters are also known).
Alternatively, if the displacement is of many wavelengths, counting minima and maxima of interference can leads to reconstruction of the displacement.
In practice though, this signal analysis is often complex and remains error prone due to detection noise, very large signal bandwidth (even for very simple displacement patterns) or variations of the reflectivity of non-cooperative targets.
Many approaches have been designed to mitigate these long-standing issues. At the physical level, tracking speckle grains or more generally ensuring a constant operating regime has been achieved by adaptive optics or liquid lenses for instance \cite{norgia2001interferometric,zabit2010adaptive,bernal2014robust}. At the software level, several techniques have also been explored to accurately reconstruct the displacement of the target from these complex signals \cite{magnani2012self,arriaga2014speckle,siddiqui2017all,usman2019detection,bernal2021toward} (including in real time for the first one). However, in spite of these progresses, data processing remains a challenging task, especially when spectrally broad displacement must be reconstructed.
As was shown recently in \cite{barland2021convolutional} a relatively small convolutional neural network (CNN) can recover the (spectrally decade-broad) displacement of a target from the voltage across the laser diode with very high precision, without the complexity of parameter tuning typical of conventional fringe counting approaches. This method has recently been extended to a multichannel measurement scheme, which results in much improved measurement availability \cite{matha2023high}. In \cite{ahmed2019self} artificial neural networks (ANNs) were used for signal pre-processing, in \cite{kou2020fringe} neural networks were used for fringe discrimination in the signal and \cite{an2022measuring} discussed the use of neural networks for parametric conditions estimation. 

From these works, it is clear that the development of machine-learning-based solutions for self-mixing displacement (or more general purpose \cite{brambilla2020versatile}) sensors is extremely promising, provided that such approaches can be integrated into a standalone instrument that includes the physical sensor (light source and detection), the data-acquisition stage, and the data-processing component.

The STMicroelectronics STM32 family of low-power 32-bit microcontrollers can offer such a device to run the inference of tiny deep neural networks.
To achieve this, a software framework that enables the deployment of deep neural networks on microcontrollers must be used.
Popular frameworks include TensorFlow Lite for Microcontrollers and STM32Cube.AI, but we chose to use Qualia\cite{qualia} for its low memory footprint and its ease of customization\cite{microaiarticle}.
The neural network is quantized to fixed-point numbers in order to keep the memory footprint low and to obtain the best performance.

Furthermore, the embedded platform can be designed so that it captures the signal directly through an ADC (Analog-to-Digital Converter), rather than requiring an external capture device to acquire the data on a workstation.
Finally, the use of a microcontroller dedicated to this task, with a high level of control over the data capture, enables real-time operation.
Therefore, an updated prototype of a self-contained device for the sensing of micrometric displacements through self-mixing interferometry is presented in this article.

Our contributions are as follows:
\begin{itemize}
  \item design of a tiny deep neural network for small displacement measurements through self-mixing interferometry,
  \item ensure the neural network operates correctly independently of the signal amplitude
  \item evaluation of the quantization of the neural network,
  \item evaluation of embedded performances after deployment on microcontroller,
  \item demonstration of realtime operation of the sensor
\end{itemize}

The present article is an extended version of a prior publication at the 2024 IEEE Sensors Applications Symposium\cite{IEEESAS2024}.
The new contributions include:
\begin{itemize}
  \item an improved deep neural network based on the ResNet architecture compared to the less efficient CNN and better training hyperparameters,
  \item a new embedded platform based on an STM32U575ZI microcontroller compared to the previous STM32L476RG,
  \item the live inference in real-time thanks to both previous improvements,
  \item a filtered digital-to-analog converter (DAC) output of the predictions,
  \item an input normalization step during inference to be more robust again variations in amplitude,
  \item and finally an evaluation of the energy consumption of the system.
\end{itemize}

The rest of this article is organized as follows.
Section \ref{sec:method} explains the deep neural network, its quantization and deployment on the microcontroller.
Section \ref{sec:setup} describes the optoelectronics setup and the embedded platform.
Section \ref{sec:results} presents results on the prediction performance of the neural network before and after quantization, as well as results on the embedded execution.
Finally, section \ref{sec:conclusion} concludes our work and provides future perspectives.

\section{Proposed method}
\label{sec:method}

\subsection{ResNet based reconstruction}
\label{subsec:resnet}
\begin{figure}[ht]
    \center
    \includegraphics[width=\linewidth]{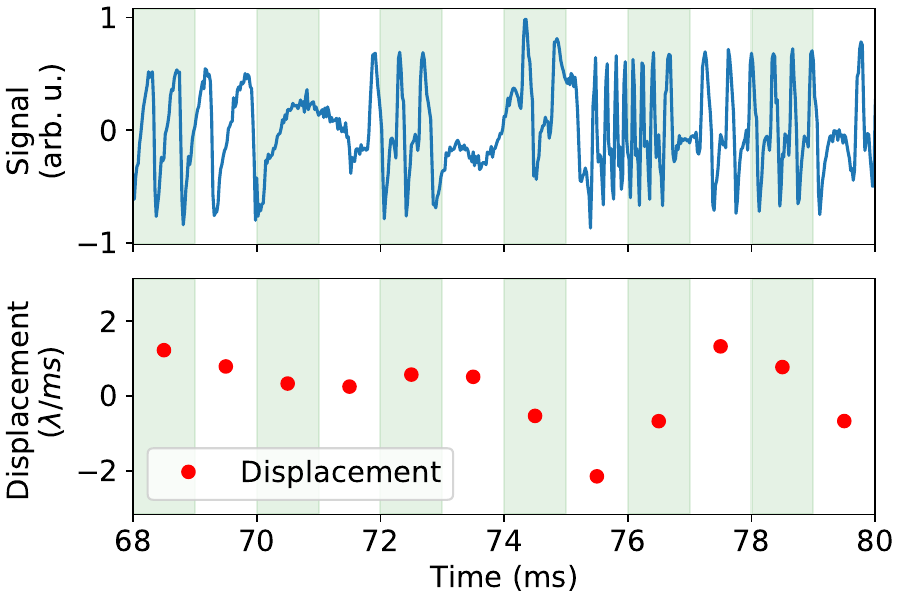}
	\caption{Example of the regression task to be realized. Top: the interferometric signal is measured as a voltage at the edges of a semiconductor laser diode in the course of time. Bottom, the displacement of the target during the same time frame. The task is to reconstruct the displacement of the target (red dots) from the signal during the corresponding time segment (the alternating light green and white stripes)}    
    \label{fig:example}
\end{figure}

The task to be realized by the network is represented on \ffig{fig:example}. On the basis of segments of the interferometric signal in the course of time, the network must infer the value of the displacement of the target during the considered time interval. One of the peculiarities of this signal is that, since it consists essentially of interference fringes taking place in a nonlinear medium, the amplitude of the signal is not significant in terms of the displacement of the target but only in terms of its reflectivity. Thus, the information which must be extracted from the signal resides in the shape of the fringes, their more or less peaked aspect, their orientation and their repetition rate.

Achieving real-time operation on a microcontroller-based platform requires meeting several constraints: the model must have a sufficiently small memory footprint to fit on the device; the number of multiply–accumulate operations per inference must remain low enough to satisfy the timing requirements (here, 1 ms per sample); and the inference error must be bounded (here, below half a wavelength per sample), regardless of the signal amplitude.

\begin{figure}[]
    \center
\includegraphics[height=0.50\textheight]{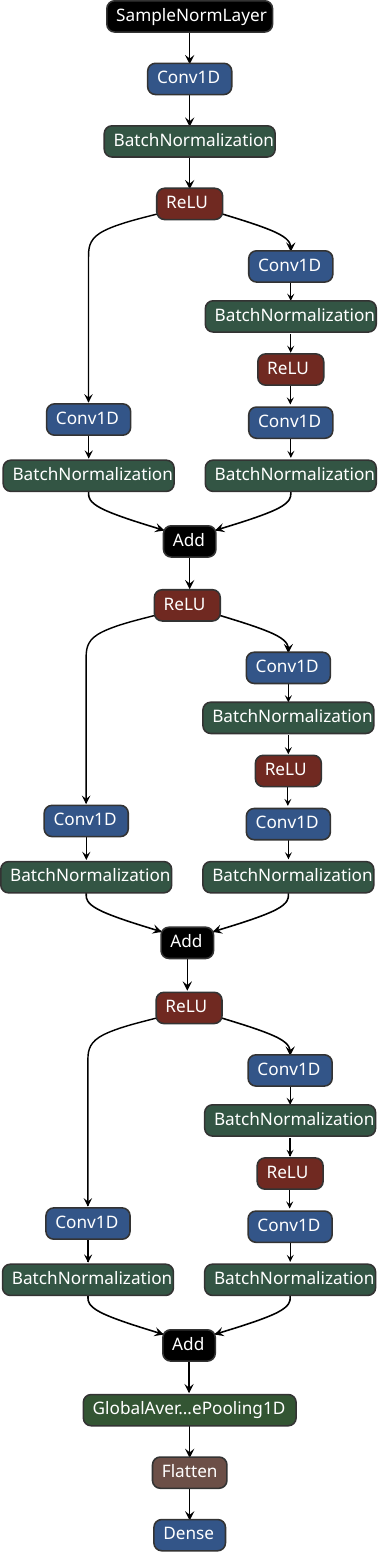}
	\caption{Neural network structure. Same length padding is used at each convolution. All kernels have length 3 except in the residual branches (length 1).}
    \label{fig:resnet}

\end{figure}

The first attempt at a basic neural network for this task was originally described in \cite{barland2021convolutional} and consisted of a an alternating stack of four convolutional and four maxpooling layers followed by two fully connected layers for the final regression.
In the original work, the network (featuring $5.7\times 10^4$ parameters) processed segments of 256 points acquired at a sampling rate of 250kS/s.
A first downscale of this network featuring 21~601 parameters was presented in \cite{IEEESAS2024}. 
In order to run the network in real time on the STM32 platform, we prepared a more efficient network based on the ResNet architecture \cite{he2016deep}, shown on \ffig{fig:resnet}.
The network (which features 6937 parameters after BatchNorm fusion) is designed to operate on non-overlapping windows of duration 1~ms with 48 samples acquired at a rate of 48kHz. We note that our architecture choice is not the only possible one and it may be that other architectures designed for sequence analysis tasks (\textit{eg} transformers) could be more efficient, but as we show below this simple approach works well. 

In addition, since the network must operate on a signal whose amplitude may vary depending on the reflectivity of the target or on the specific model of laser used at the physical layer, we have inserted a normalization layer at the input of the network which scales each input segment between 0 and 1.
Although this layer is not trainable and could therefore be considered a preprocessing task which does not belong to the network, inserting it into the network itself proved convenient.

\subsection{Network training}
\label{subsec:training}
To optimize the network for the reconstruction task, we start from the data set available in \cite{barland2022displacement}. As described in \cite{barland2022displacement,barland2021convolutional}, this dataset consists of two subsets of experimental data, one for network design and preparation phase (training/validation), and another one for a final generalization performance assessment. The first subset consists only of periodic displacements over a 5-100~Hz frequency range and -2.5 to +2.5 $\lambda/ms$ amplitude of displacements, including many points near 0 displacement (195~011 segments and their corresponding displacement). As explained in \cite{barland2021convolutional}, this experimental dataset is built on many different feedback alignment conditions and $C$ parameter values. The out-of-distribution final generalization dataset (also experimental) consists of 3438 segments with their corresponding true displacement which are aperiodic displacements covering the same amplitude and frequency ranges as the training/validation sets, but with again a different feedback alignment condition and C value. The ground truth for the displacement in this dataset comes from a calibration of the $\lambda/V$ scaling factor or a speaker which is linear (within 10\%) over the selected operation frequency.

This data set was downsampled so that each segment of signal consists of 48 points instead of 256.
In addition, the original data set is normalized by its standard deviation.

The network is trained during 120 epochs using the Adam optimizer with an initial learning rate of 0.002.
The learning rate is divided by 3 at epochs 20, 50, 70, 85, 100, 110 and 115.
A mean-squared error (MSE) loss function is used and a weight decay of $5\times 10^{-4}$ is applied. As usual for neural networks training, these parameters influence the speed and quality of convergence and the number of epochs was chosen to reach the global minimum of the loss function. Tuning these values relies on expertise in neural networks training and direct experimentation and we refer the reader to textbooks \cite{goodfellow2016deep} for standard approaches.
Random noise following a normal distribution of mean 0 and standard deviation of 0.15,
as well as random cutouts with lengths following a normal distribution of mean 0 and a standard deviation of 0.2 are used as data augmentation techniques. 

The CNN presented in \cite{IEEESAS2024} has also been re-trained with these settings to compare the results with the ResNet presented in section \ref{subsec:resnet}.

The data used for training procedure is the periodic motion subset and the aperiodic subset is used for generalization evaluation discussed in section \ref{sec:results}.

\subsection{Artificial neural network quantization}
After training the artificial neural network, the inputs, the weights and the activations are quantized.
This enables a reduction both of memory footprint and latency.
Indeed, going from 32-bit floating-point numbers to 16-bit fixed-point numbers halves the memory usage of the weights (in ROM), the activations (in RAM) and the inputs (in RAM).
Furthermore, computation with fixed-point numbers can be better optimized to be performed faster than floating-point numbers.

The quantization of the neural network follows the method presented in \cite{microaiarticle}.
Post-training quantization with nearest rounding mode, a power-of-two scale factor and a symmetric range is used.
The same scale factor is applied to weights, biases, activations and inputs.
For 16-bit quantization, the empirically chosen format is Q7.9 for the entire network, where Qx.y means x bits for the integer part (including sign) and y bits for the fractional part.
For 8-bit quantization, the format is selected per-layer and independently for weights and activations.
In this case, the number of bits for the integer part is chosen automatically to be able to represent the largest values encountered during training.
The remaining bits are allocated to the fractional part.

\subsection{Deployment on microcontroller}
Prior to deployment, the batch normalization layers are fused with the preceeding convolutional layer.

The Qualia framework\cite{qualia} is used in order to generate a portable C library for inference from the trained model.
An inference function for each layer of the neural network along with the layers call chain are generated.

In order to speed the inference up, we use the CMSIS-NN\footnote{\textsc{CMSIS} stands for Common Microcontroller Software Interface Standard}\cite{CMSISNN} library.
CMSIS-NN contains optimizations for computation of neural network operations such as convolutions, and makes use of specific instructions when available
(e.g. SMLAD to perform two 16-bit integer multiply-accumulate operations in a single cycle).

The generated C code is then compiled with the GCC compiler, using the \texttt{-Ofast} optimization level and LTO (Link-Time Optimization), and uploaded to the microcontroller.

\section{Experimental setup}
\label{sec:setup}

\begin{figure}[ht]
  \includegraphics[width=\linewidth]{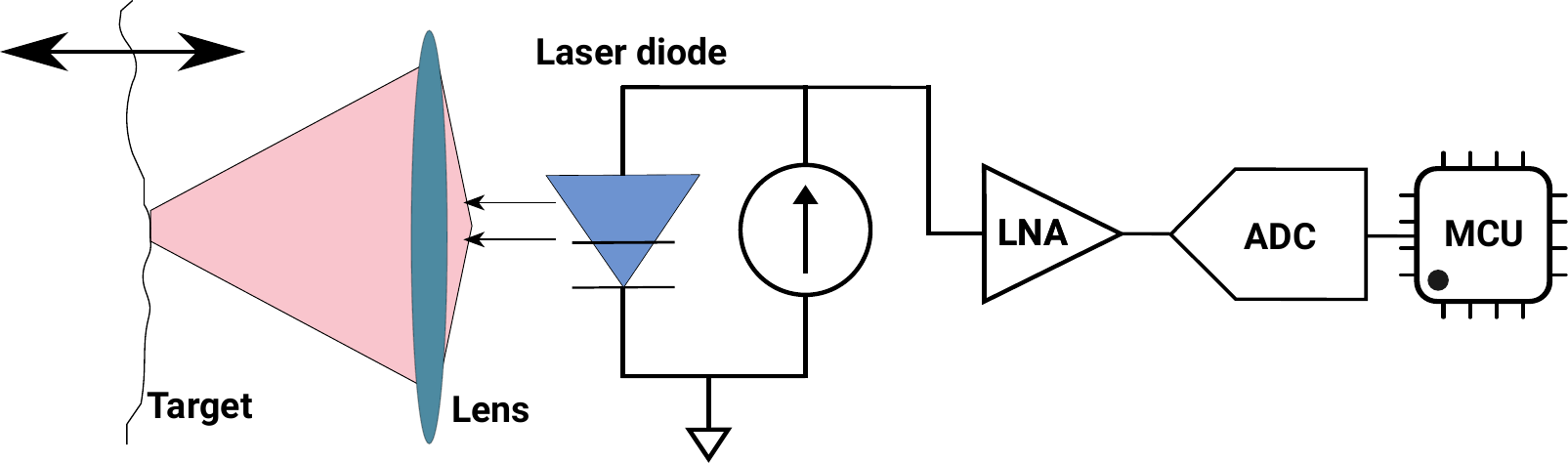}
  \caption{Experimental setup, LNA: Low-Noise Amplifier, ADC: Analog-to-Digital Converter, MCU: MicroController Unit}
  \label{fig:setup}
\end{figure}

\subsection{Optoelectronics setup}
The experimental arrangement is shown schematically on \ffig{fig:setup}.
From the optical point of view it consists of a semiconductor laser driven slightly above its coherent emission threshold by a stabilized current source emitting a beam of about 1~$mW$ power at a wavelength $\lambda=1.310~\mu m$.
The beam is focused by a lens on a target at a distance of about 15~cm and a tiny part of it (of the order of $10^{-4}$) is re-injected back into the laser.
In this case the target is a simple speaker whose displacement has been calibrated so that the voltage at the edges of the speaker can be used as a proxy for the displacement of the target.
The details of this optical system are available in \cite{barland2021convolutional} and we omit them here for brevity.
The interferometric signal is obtained by measuring the voltage at the edges of the laser diode.
This voltage consists of a large (about 1~V) continuous component superimposed with much smaller fluctuations (sub-mV) which contain the relevant part of the signal.

An external AC-coupled low-noise amplifier built using a pair of LT1128 with a total gain of approximately $10^4$ is placed between the laser and the embedded platform.
The low-noise amplifier circuit is fed with a $+15V/-15V$ symmetric supply, however its output is clamped in order to prevent exceeding the ADC's 3.6~V maximum rating.
The amplitude of the signal at the output of the low-noise amplifier varies depending on the calibration of the optoelectronics setup, however a good calibration provides between 400~mV and 1.2~V peak-to-peak.

\subsection{Embedded platforms}
\begin{figure}[ht]
    \center
    \includegraphics[width=0.7\linewidth]{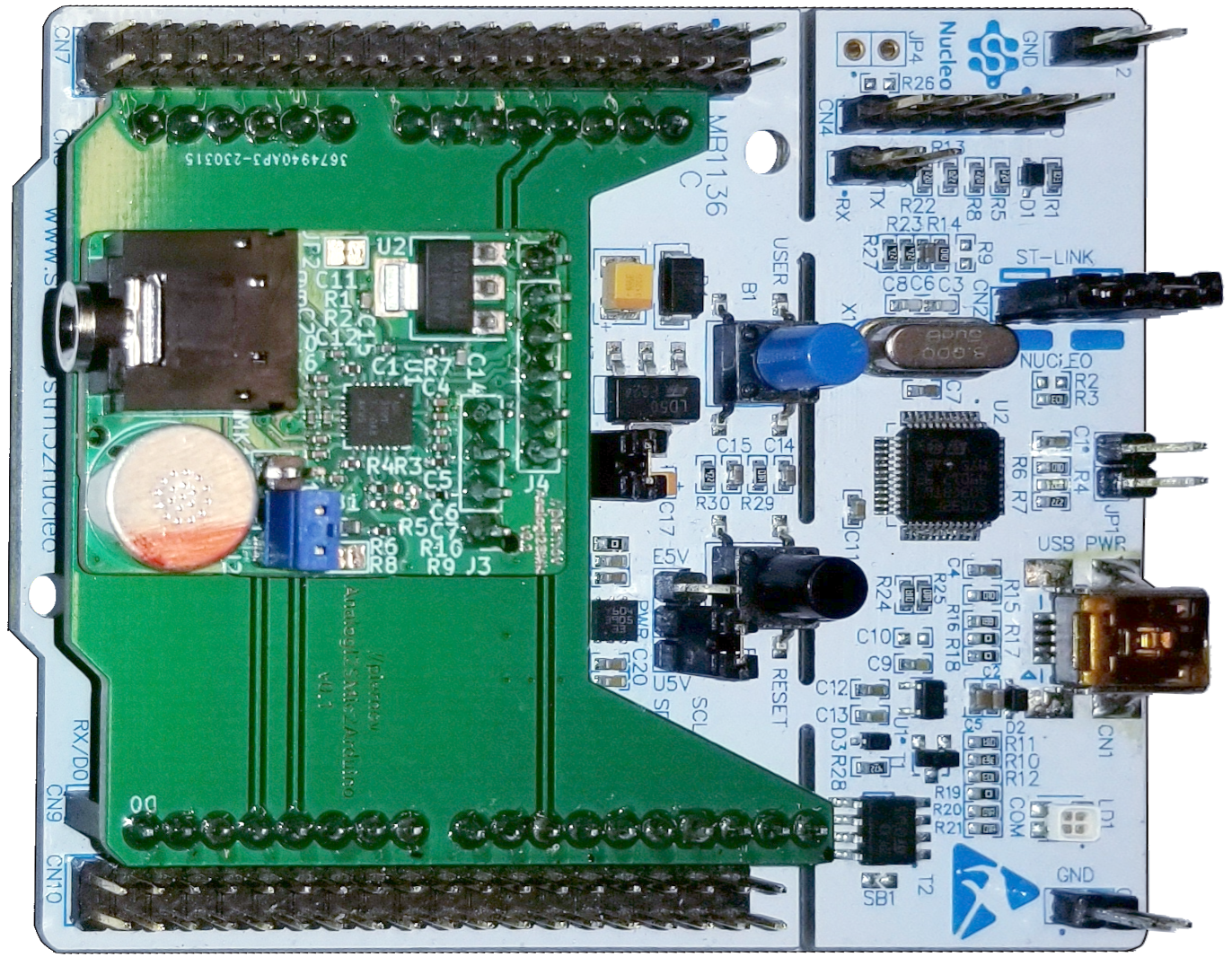}
    \includegraphics[width=0.7\linewidth,angle=180.75]{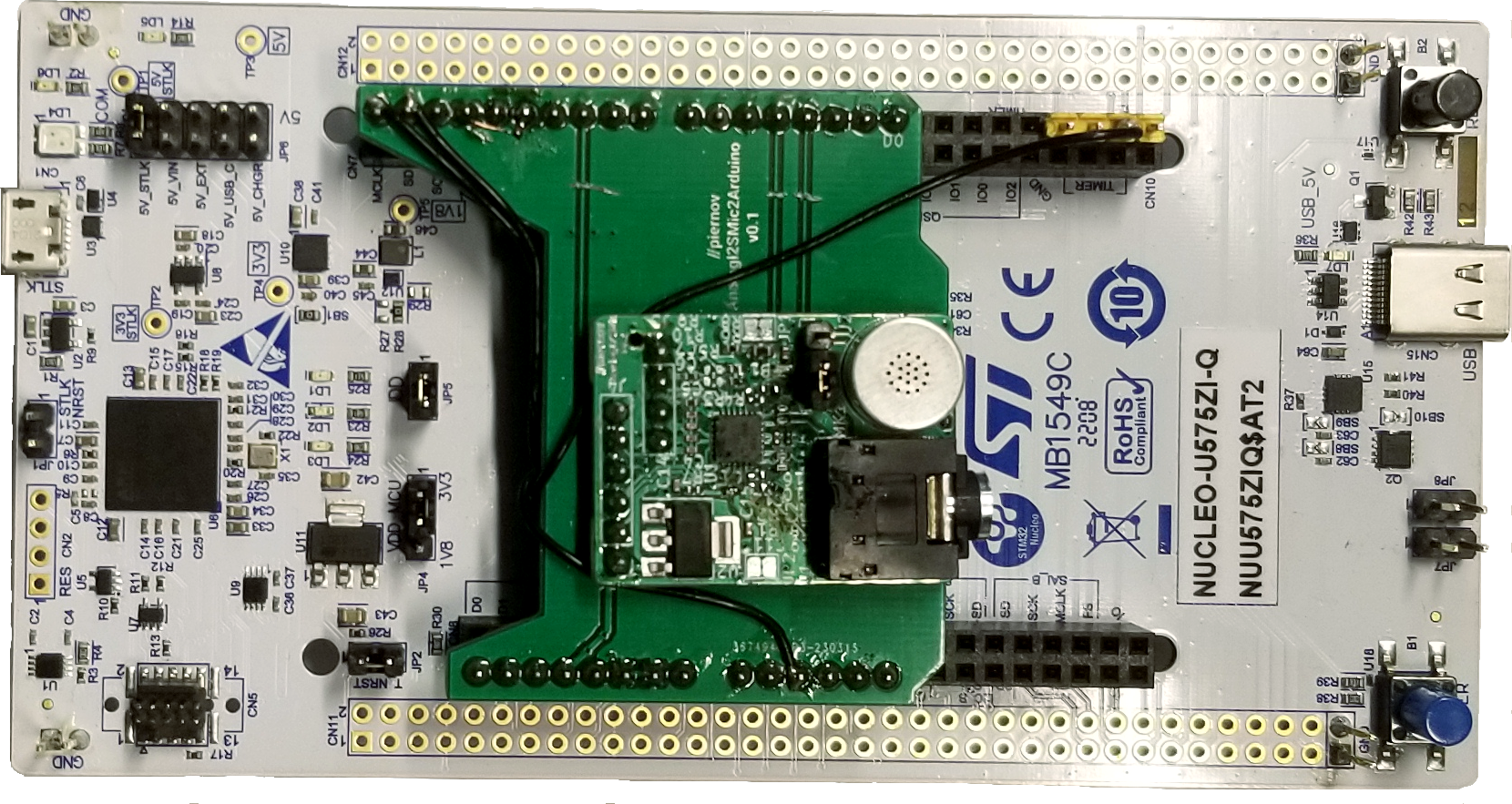}
	\caption{Top: Nucleo-L476RG embedded platform, bottom:Nucleo-U575ZI-Q embedded platform. Both boards are equipped with a custom capture board based on a TLV320ADC3101 Analog to Digital Converter (see text).}    
    \label{fig:embedded_platform}
\end{figure}

Our prototyping embedded platform consists of an STMicroelectronics Nucleo microcontroller board (white PCB) mounted with a custom capture interface (green PCB).

Two different microcontroller-based development boards are compared (Fig. \ref{fig:embedded_platform}):
\begin{itemize}
	\item the Nucleo-L476RG board , designed around an STM32L476RG microcontroller with a Cortex-M4F core running at 80~MHz, 1~MiB of Flash memory and 128~KiB of SRAM \cite{rg},
  \item the Nucleo-U575ZI-Q board, designed around an STM32U575ZI microcontroller with a Cortex-M33 core running at 160~MHz, 2~MiB of Flash memory and 768~KiB of SRAM \cite{zi}.
\end{itemize}

The custom capture interface is originally designed for audio applications with microphones.
It can also be used to capture other AC signals, although the TLV320ADC3101 we used as an analog-to-digital converter only supports standard audio sampling rates.
The TLV320ADC3101 is configured with unity gain to the input signal, a sampling rate of 48~kHz, and a resolution of 16 bits.
Only one channel is used.
The signal is fed through a 3.5~mm jack connector.

After being captured by the ADC, the signal is sent to the microcontroller through an I2S (Inter-IC Sound) bus, and a DMA (Direct Memory Access) stores the data into a buffer in RAM.
The buffer size corresponds to the size of the input window that the neural network is designed to process.
A double-buffering system is implemented so that the data can continue to be collected into one buffer while the other buffer is being processed by the artificial neural network.

For evaluation of the embedded platform using data from the test dataset, the output of the low-noise amplifier is replaced with the audio output of a laptop which streams the interferometric signal from the test dataset.
For live reconstruction, the output of the neural network is sent to a digital-to-analog converter (integrated into the microcontroller) and then filtered with a passive first order low-pass RC filter with a cutoff frequency of 240~Hz.

\section{Results}
\label{sec:results}

\subsection{Robustness to amplitude variations on workstation}
\label{subsec:robust}

As outlined in \ref{sec:method}, the network must be able to reconstruct the displacement of the target even in presence of unwanted variations of the signal amplitude. Two methods have been evaluated. The first one consists in training the network with random variations of the input in order to make it insensitive to these signal variations. The other approach is to scale each sample at the input so that the remaining part of the network processes perfectly scaled date. We therefore assess the impact of variations of the input signal amplitude with two networks which are identical except for the scaling layer inserted on top of one of them. We note that scaling input data (especially for multiple inputs of different magnitude and dimensions) is a usual preprocessing task for neural networks training and operation. However, our real-time operation goal implies that we do not have access to any statistical information to scale our input data. Instead, we must deal with (potentially large) sample to sample variations. An additional benefit is that the adequately prepared neural network will be able to operate with different laser systems, adapting to any signal amplitude provided by the upstream Analog to Digital converter.

We underline that all results presented here are \textit{generalization} results, which are obtained using the (out of ditribution) aperiodic displacement subset of \cite{barland2022displacement}, while all training was done with only the periodic subset (split into 80\% for training and 10\% each for validation and test).

Here, to simulate spurious signal amplitude variations, we apply during training (and during the generalization evaluation shown below) a random multiplication factor comprised between 1 and 100 to each segment of input signal. Both networks are trained during 40 epochs with the RMPSprop optimizer and an initial learning rate of $10^{-3}$ which is decreased progressively down to $6\times 10^{-5}$ on learning plateaus, 20\% of the data set being kept for validation. In order to compare the performance between the two models (with and without input scaling layer), all trainings have been realized deterministically by seeding identical initializers and data shuffling. Of course, the random scaling during training has no impact on the network which includes a normalization layer.

\begin{figure}[ht]
    \center
    \includegraphics[width=\linewidth]{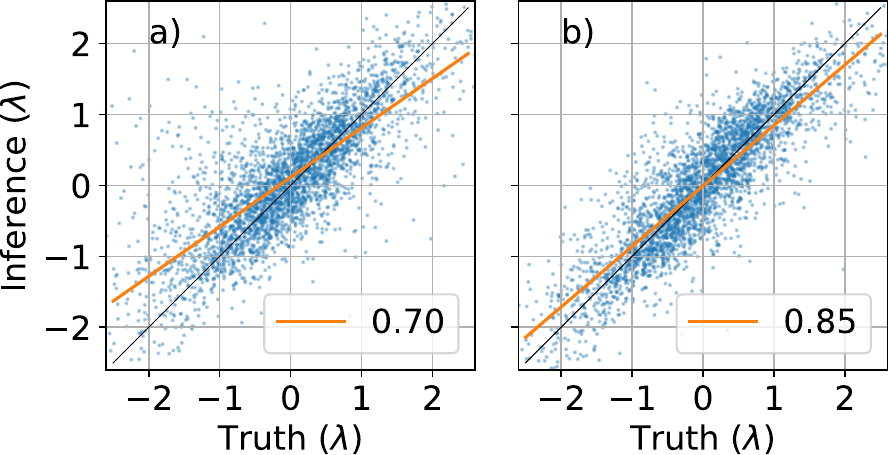}
	\caption{Performance comparison between a network without input normalization layer (a) and with input normalization layer (b) on 3438 measurement points from the generalization set. The legends indicate the linear regression coefficient between true and inferred displacement for each model.}    
    \label{fig:scaling1}
\end{figure}

The results of generalization inference for both models are shown on \ffig{fig:scaling1}. On both panels we show the inferred displacement against the true displacement using the aperiodic displacement signal provided in \cite{barland2022displacement} after a random scaling between 1 and 100 has been applied to each segment (as was done for the training). A perfect model would of course show a perfect straight line with slope 1 and intercept 0. On the left panel, we observe that the model which does not include a scaling layer behaves rather poorly, with a strong under-estimation of the displacement and large dispersion of the inferred values (Mean Absolute Error MAE=0.43, Mean Squared Error MSE=0.40). With the scaling layer, the results are much improved, with a linear regression coefficient of 0.85 (MAE=0.35, MSE=0.22). We underline that this generalization test set is markedly different from the training (and validation/test) sets, which ensures that the sensor can operate in a variety of experimental conditions. In particular for the smallest displacements (below 0.1~$\lambda$) the inferred value has a standard deviation of 0.42~$\lambda$, which emphasizes the stability of the system near zero displacement (discussed in depth in \cite{barland2021convolutional}). If we realize the same performance assessment using the (in distribution) validation and test data subsets, which were not used for training but are much more similar to the training data since they are all periodic displacements, the slope obtained by the model with scaling layer is 0.98 for the ($3.9*10^4$ samples) validation set (MAE=0.14, MSE=0.04) and 0.96 (MAE=0.137, MSE=0.429) for the test set. We stress however that for this application, the key result lies in the out of distribution generalization test realized on the aperiodic motion subset. 

\begin{figure}[ht]
    \center
    \includegraphics[width=\linewidth]{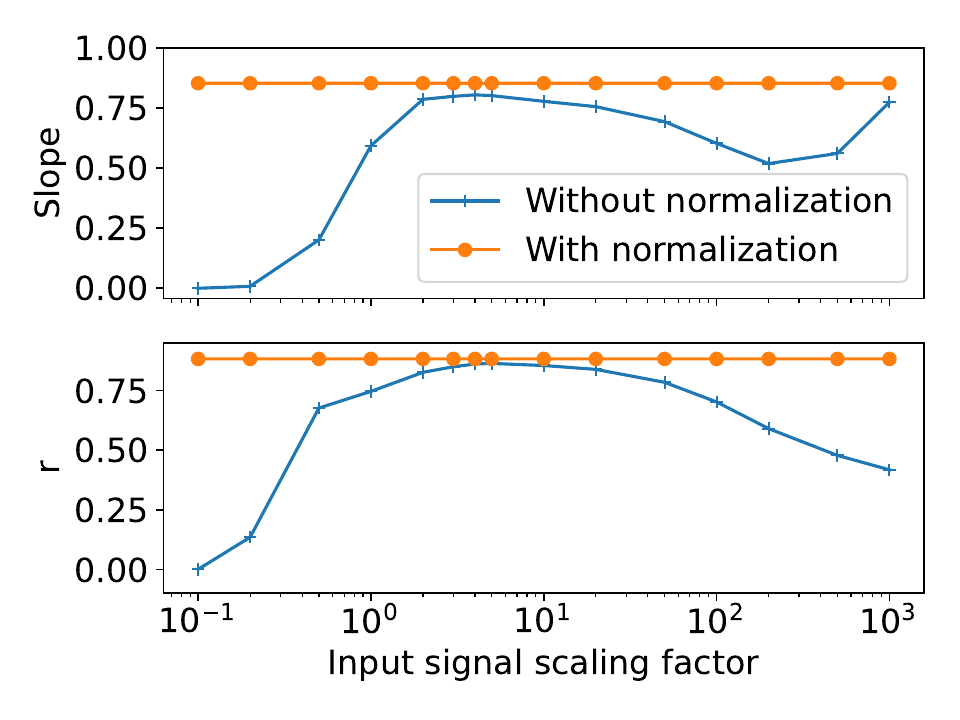}
	\caption{Robustness against input signal amplification for models with and without input normalization layer. Top: slope of the linear regression coefficient. Bottom: Pearson's $r$ correlation coefficient.}    
    \label{fig:scaling2}
\end{figure}

At this point, we have shown that the scaling layer ensures a globally correct inference in presence of random fluctuations of the input amplitude. We further compare the two models on \ffig{fig:scaling2}, where we show how the linear regression slope and the Pearson's correlation coefficient $r$ vary with the input signal amplification. 
We observe that applying random amplifications during training does have a positive impact since the model without normalization shows best performance within the range in which the training was performed. 
The performance is however far from constant even if one where to consider only a narrow range of operation of a single order of magnitude for the input signal. 
Instead, the normalization layer allows for constant performance operation over four orders of magnitude, which ensures robust operation and relaxes the constraint on the specifications of the low-noise amplifier.

\subsection{Evaluation of quantization}

The effect of quantization of the neural network is evaluated by generating the C code with Qualia.
Three configurations are evaluated: floating-point numbers with no quantization (float32), fixed-point 16-bit quantization (int16), and fixed-point 8-bit quantization (int8).
The inference is then performed on a workstation with the test dataset.

\begin{figure}[ht]
  \center
    \includegraphics[width=0.7\linewidth]{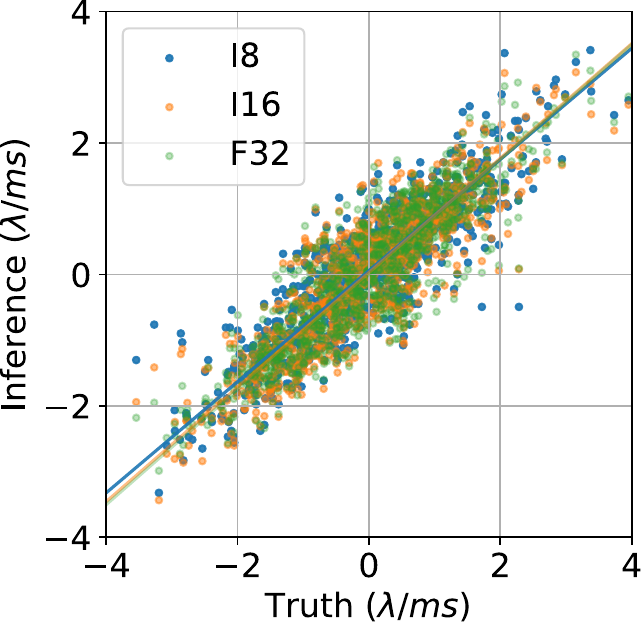}
  \caption{Predictions for float32 (F32), int16 (I16) and int8 (I8) models vs. ground truth. The solid lines are optimal linear fits with coefficients 0.87, 0.87 and 0.84 for float32, int16 and int8 respectively. }
    \label{fig:quantization_dispersion}
\end{figure}

\begin{table}[ht]
  \centering
  \caption{\textsc{Pearson Correlation Coefficient (PCC), Linear Regression Slope (LRS), Mean Squared Error (MSE) and Mean Absolute Error (MAE) for Float32, Int16 and Int8 Predictions Compared to Ground Truth}}
    \begin{tabular}{|c|c|c|c|r|r|r|r|r|}
      \hline
      \textbf{ANN model} & \textbf{Data format} & \textbf{PCC} & \textbf{LRS} & \textbf{MSE} & \textbf{MAE} \\
      \hline
      CNN\cite{IEEESAS2024} & float32 & 0.8551 & 0.74\phantom{00} & & \\
      \hline
      CNN & float32 & 0.8624 & 0.7964 & 0.258 & 0.388 \\ %
      \hline
      CNN & int16 & 0.8521 & 0.7039 & 0.273 & 0.388 \\ %
      \hline
      CNN & int8 & 0.8479 & 0.6701 & 0.288 & 0.399 \\ %
      \hline
      ResNet & float32 & 0.8843 & 0.8754 & 0.227 & 0.364\\
      \hline
      ResNet & int16 & 0.8893 & 0.8714 & 0.216 & 0.353\\
      \hline
      ResNet & int8 & 0.8873 & 0.8461 & 0.219 & 0.353\\
      \hline
    \end{tabular}
  \label{tab:quantization_correlation}
\end{table}

Fig. \ref{fig:quantization_dispersion} shows the scatter plot for predictions of the neural network for int8 quantization (blue), int16 quantization (orange), and non-quantized float32 (green) versus the ground truth. No obvious staircase effect is visible for any of the quantizations but it is worth underlining that quantization takes place at all depths in the network (and not only the output layer) so the effects of quantization must be evaluated more quantitatively, which we discuss in the following.
Table \ref{tab:quantization_correlation} provides the Pearson correlation coefficient and the slope of the linear regression with respect to the ground truth for each configuration.

The ResNet provides significantly better correlation coefficient and linear regression slope than the previously-designed CNN.
For the ResNet, the difference in correlation coefficient and slope between float32 and int16 are respectively $+0.005$ and $-0.004$.
We consider this not to be significant.

On the other hand, int8 quantization causes a loss in prediction performance as the slope drops by 0.0293 compared to float32.
This highlights the limits of such a simple quantization scheme.
The drop in prediction performance is even more significant for the CNN, meaning that the ResNet provides better prediction quality even at low bitwidth.
In order to improve the results of 8-bit quantization,
a better quantization scheme with asymmetric quantization for the activations and a separate scale factor for the bias could be used.
Quantization-aware training could also be considered.\cite{QualcommQuant}

\subsection{Evaluation of embedded constraints}

\begin{table}[ht]
  \centering
  \caption{\textsc{Latency and Memory Consumption on STM32L476RG for Float32, Int16 and Int8 Inference, with and without CMSIS-NN}}
  \resizebox{\linewidth}{!}{
    \begin{tabular}{|c|c|c|c|c|c|}
      \hline
      \textbf{ANN model} & \textbf{Data format} & \textbf{CMSIS-NN} & \textbf{Latency} & \textbf{ROM} & \textbf{RAM}\\
                         &                      &                   & \textbf{(ms)}         & \textbf{(kB)} & \textbf{(kB)}\\
      \hline
      CNN & float32 & No & 15.32 & 141.800 & 26.312\\ %
      \hline
      CNN & int16 & No & 12.28 & \phantom{0}99.096 & 24.008\\ %
      \hline
      CNN & int16 & Yes & \phantom{0}4.68 & 100.056 & 25.368\\ %
      \hline
      CNN & int8 & No & 13.48 & \phantom{0}77.936 & 22.496\\ %
      \hline
      CNN & int8 & Yes & \phantom{0}3.80 & \phantom{0}81.056 & 24.024\\ %
      \hline
      ResNet & float32 & No & \phantom{0}2.91 & \phantom{0}90.152 & 22.344\\
      \hline
      ResNet & int16 & No & \phantom{0}2.38 & \phantom{0}75.824 & 21.672\\
      \hline
      ResNet & int16 & Yes & \phantom{0}1.62 & \phantom{0}73.392 & 22.808\\
      \hline
      ResNet & int8 & No & \phantom{0}2.50 & \phantom{0}69.672 & 21.336 \\
      \hline
      ResNet & int8 & Yes & \phantom{0}1.42 & \phantom{0}64.376 & 22.536 \\
      \hline
    \end{tabular}
  }
  \label{tab:embedded_measurements}
\end{table}

\begin{table}[ht]
  \centering
  \caption{\textsc{Latency and Memory Consumption on STM32U575ZI for Float32, Int16 and Int8 Inference, with and without CMSIS-NN}}
  \resizebox{\linewidth}{!}{
    \begin{tabular}{|c|c|c|c|c|c|}
      \hline
      \textbf{ANN model} & \textbf{Data format} & \textbf{CMSIS-NN} & \textbf{Latency} & \textbf{ROM} & \textbf{RAM}\\
                         &                      &                    & \textbf{(ms)}         & \textbf{(kB)} & \textbf{(kB)}\\
      \hline
      CNN & float32 & No & 5.97 & 144.632 & 26.416\\ %
      \hline
      CNN & int16 & No & 5.12 & 101.992 & 24.112\\ %
      \hline
      CNN & int16 & Yes & 2.03 & 102.984 & 25.472\\ %
      \hline
      CNN & int8 & No & 5.97 & \phantom{0}80.880 & 22.608\\ %
      \hline
      CNN & int8 & Yes & 1.82 & \phantom{0}85.200 & 24.128\\ %
      \hline
      ResNet & float32 & No & 1.09 & \phantom{0}93.040 & 22.448\\
      \hline
      ResNet & int16 & No & 1.05 & \phantom{0}78.696 & 21.776\\
      \hline
      ResNet & int16 & Yes & 0.64 & \phantom{0}76.376 & 22.912\\
      \hline
      ResNet & int8 & No & 1.15 & \phantom{0}72.520 & 21.440 \\
      \hline
      ResNet & int8 & Yes & 0.62 & \phantom{0}67.288 & 22.640 \\
      \hline
    \end{tabular}
  }
  \label{tab:embedded_measurements_u5}
\end{table}

Table \ref{tab:embedded_measurements} shows latency and memory for different configurations of the neural network when deployed on the STM32L476RG microcontroller.

The latency is measured as the time taken to perform the inference, after collecting a window of data in the input buffer.
Both ROM and RAM footprints are provided. They are extracted from the memory allocation in the firmware after compilation.
The ROM contains the program instructions and the neural network weights.
The RAM contains the input buffer and each layer's output buffer, all allocated statically.
When possible, i.e., on independent layers, buffers share a common memory space.
Note that some memory is also used for other parts of the firmware, e.g., serial communications.

In these results, we compare the effect of fixed-point quantization, with and without CMSIS-NN optimizations.
Compared to 32-bit floating point, 16-bit quantization reduces the total ROM footprint by approximately 30\% for the CNN and 21\% for the ResNet.
Without CMSIS-NN, the reduction in inference time after 16-bit quantization is between 5\% and 20\%.
However, using CMSIS-NN allows for a decrease of close to 70\% for the CNN and 45\% for the ResNet.

8-bit goes further in terms of memory footprint reduction, by approximately 45\% for the CNN and 29\% when compared to 32-bit floating point.
Inference with 8-bit fixed point is slightly faster than with 16-bit fixed point when CMSIS-NN is used, providing up to 75\% reduction in latency for the CNN compared to 32-bit floating point.

All of these configurations fit well within the ROM and RAM constraints (1024~kiB and 128~kiB, respectively), however the latency is well above our real-time constraint of 1~ms.

Table \ref{tab:embedded_measurements_u5} shows the same measurements for the STM32U575ZI microcontroller.
All these configurations also fit well within the memory limits (ROM limit: 2048~kiB, RAM limit: 768~kiB).

As can be seen in the latency figures, this microcontroller is two to three times faster, attributed to the higher core clock and the more advanced architecture.
The CNN still cannot reach the real-time computation constraint on this microcontroller.
However, the more efficient ResNet is able to perform the computation in less than 1~ms when quantized in fixed-point and using the CMSIS-NN library.
Therefore, this configuration is able to process the interferometric signal in real time.

The improvement provided by 8-bit quantization is not significant enough when compared to the 16-bit quantization to justify bearing the lower prediction performance.
Furthermore, we already concluded in the previous section that 16-bit quantization does not significantly deteriorate the quality of the predictions over 32-bit floating point.
Therefore, our choice is to use the 16-bit quantized ResNet with CMSIS-NN running on the STM32U575ZI microcontroller for real-time reconstruction.

\subsection{Evaluation of energy consumption}

\subsubsection{Laser diode}
The laser is fed with an ajustable constant-current power supply.
The best interferometric signal is typically observed between 6.5~mA and 8.5~mA.
In this experiment, the current is set to 7.2~mA.
Using a LeCroy WaveSurfer 104Xs oscilloscope, the average DC-coupled voltage is measured to be 1.0~V in this configuration, amounting to an average power consumption of 7.2~mW.

\subsubsection{Low-noise amplifier}
The power consumption of the low-noise amplifier is measured with a Brymen BM869s in mA range.
The voltage between the positive and negative supply rails is measured to be 29.2~V, with a current of 16.5~mA, amounting to a power consumption of 481.8~mW.

\subsubsection{Analog-to-digital converter}
The average power consumption of the analog-to-digital conversion board is measured with a Qoitech Otii Arc power analyzer, set to a constant voltage of 3.3~V.
A typical 1.1~V peak-to-peak interferometric signal is injected at the input of the ADC.
In this configuration, the average current consumption is 10.0~mA, giving an average power consumption of 33.0~mW.

\subsubsection{Microcontroller}

The microcontroller current consumption is measured with a Qoitech Otii Arc power analyzer set to 3.3~V in place of the MCU IDD jumper (JP2 on Nucleo-L476RG, JP5 on Nucleo-U575ZI-Q).
The microcontrollers are configured to run at their maximum rated core clock (80~MHz for the STM32L476RG and 160~MHz for the STM32U575ZI).
Since the STM32L476RG cannot perform real-time processing due to its insufficient computing performance, a 500~ms input buffer is used.
The power consumption is averaged over the data collection and inference phase for the 500~ms window and the results are listed in Table \ref{tab:mcu_energy_consumption}.
The ANN models used here are quantized on 16 bits and CMSIS-NN is used.

It is remarkable that a four-fold optimization in energy consumption is achieved by acting first on the hardware side and then on the software side. First, the best-performing platform carries out the same inference at half the energy cost of the other platform, thanks to only marginally higher power consumption but significantly faster computation. Second, on this optimal platform, the ResNet architecture provides an additional two-fold reduction in energy consumption, due to a computation time that is twice as fast while maintaining nearly constant power consumption.

\begin{table}[ht]
  \centering
  \caption{\textsc{Microcontroller Power and Energy Consumption for 500~ms of Input Signal}}
  \resizebox{\linewidth}{!}{
    \begin{tabular}{|c|c|c|c|c|c|}
      \hline
      \textbf{Microcontroller} & \textbf{ANN model} & \textbf{Total time } & \textbf{Average power } & \textbf{Energy }\\
                               &                    & \textbf{(ms)}         & \textbf{(mW)} & \textbf{($\mu$Wh)}\\
      \hline
      STM32L476RG & CNN & 3957 & 62.4 & 68.6 \\
      \hline
      STM32U575ZI & CNN & 1676 & 79.5 & 37.0 \\
      \hline
      STM32U575ZI & ResNet & \phantom{0}797 & 79.2 & 17.5 \\
      \hline
    \end{tabular}
  }
  \label{tab:mcu_energy_consumption}
\end{table}

\subsubsection{Total energy consumption for one prediction window}

For the total energy consumption figures in Table \ref{tab:total_energy_consumption}, it is assumed that the laser, LNA and ADC stay active all the time, since the purpose is to perform real-time processing where no data is missed.
Furthermore, the results are given for 1~ms which corresponds to the prediction period.
Since only the STM32U575ZI microcontroller with the ResNet neural network can perform the processing fast enough, the energy figure cannot be given for the other configurations.

\begin{table}[ht]
  \centering
  \caption{\textsc{Total Energy Consumption for 1~ms of Input Signal}}
  \resizebox{\linewidth}{!}{
    \begin{tabular}{|c|c|c|c|c|c|c|c|}
      \hline
      \textbf{Microcontroller} & \textbf{ANN model} & \multicolumn{5}{|c|}{\textbf{Average power}} & \textbf{Energy}\\
				& 			&\multicolumn{5}{|c|}{		\textbf{(mW)}} & \textbf{($\mu$Wh)}\\
      \hline
                               &                    & \textbf{Laser} & \textbf{LNA} & \textbf{ADC} & \textbf{MCU} & \textbf{Total} & \\ 
      \hline
      STM32L476RG & CNN & 7.2 & 481.8 & 33.0 & 62.4 & 584.4 & N/A \\
      \hline
      STM32U575ZI & CNN & 7.2 & 481.8 & 33.0 & 79.5 & 601.5 & N/A\\
      \hline
      STM32U575ZI & ResNet & 7.2 & 481.8 & 33.0 & 79.2 & 601.2 & 0.167\\
      \hline
    \end{tabular}
  }
  \label{tab:total_energy_consumption}
\end{table}

Overall, the highest consuming part of the setup is the low-noise amplifier.
It is important to note that it has not been optimized at all for low-power consumption.
Nonetheless, the total power consumption is around 600~mW, enabling the device to work from battery for hours.
For example, a couple of AA-sized Alkaline batteries should provide in excess of 6~000~mWh of energy, powering the device for 10 hours.
Therefore, it would be possible to build a portable autonomous sensor, completely isolated from the mains, which can be helpful in reducing the noise.

\subsection{Live inference}

\begin{figure}
   \includegraphics[width=\linewidth]{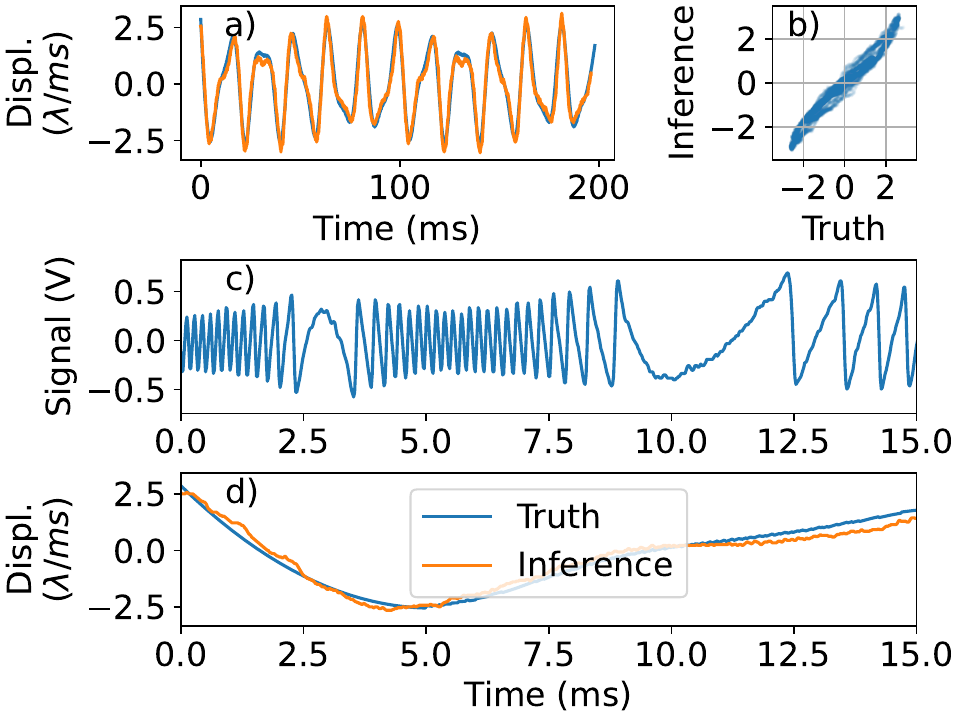}
	\caption{Example of real-time inference on the STM32U575ZI platform. a) and b): comparison between true displacement and inference over 200~ms. c) and d), signal and displacements over the first 15~ms of the time series shown in a).}
\label{fig:live_inference}
\end{figure}

In \cite{IEEESAS2024} we presented the results of onboard inference with a buffering of 500~ms due to the too long inference time which did not allow for full real time operation. 
On the contrary, with the model optimizations and the platform upgrade described above, we can show the results of fully real-time live inference on the STM32U575ZI platform on \ffig{fig:live_inference}. 
In this case, we apply to the target speaker a frequency modulated signal with frequency 70~Hz, modulation frequency 60~Hz and modulation depth 30~Hz. 
On the top left panel a), we show the true displacement (blue) and the inferred displacement (orange) over 200~ms.
The true displacement is computed as the time derivative of the instantaneous voltage applied to the speaker and scaled to units of $\lambda/ms$ by applying an independently determined $\lambda/V$ scaling factor. We have checked the linearity (within 10\%) of this signal within the frequency range 10-100~Hz.
The inferred displacement is measured from the DAC of the STM32U575ZI platform after going through the RC filter, again applying the known $\lambda/V$ conversion factor.
A time delay of 3.9~ms between the truth and the inference signals (computed by measuring the cross correlation between these signals) has been applied to compensate for data capture time, the inference time (sub-ms) and the response of the RC filter ($\simeq 0.6$~ms).
On the top right panel b), we show the inference as function of the truth. A perfect inference would be a perfect straight line of slope unity. In this particular case the quality of the reconstruction is obviously excellent. It is worthy to remark that the RC filter has an important visual impact since it plays the role of interpolating between the (downsampled) points inferred by the network. A final version of the sensor could include a more involved design of this element (here the simplest first order low-pass filter, based on a resistor and a capacitor) or simply letting the operator perform the interpolation on the basis of the step-like inference signal since these inference values are the one that the instrument can faithfully provide (the interpolation being nothing more than an interpolation...)
On bottom panels c) and d) we show for completeness a zoom over the first 15~ms of the signal and the corresponding true and inferred displacements.

We underline that this constitutes the first demonstration of a real-time displacement reconstruction based on self-mixing interferometric signal and embedded neural network.

\subsection{Discussion}

Although the main focus of the present paper is to demonstrate real time operation by embedding an optimized neural network (since offline trajectory reconstruction is thoroughly documented in \cite{barland2021convolutional,matha2023high}), a discussion of the advantages of neural networks with respect to existing alternatives appears to be useful. 

Identification of the type of fringe in the feedback signal as in \cite{usman2019detection} or estimation of the $\alpha$ parameter as in \cite{bernal2021toward} are not required if direct displacement inference is done by a neural network with adequate training data \cite{barland2021convolutional,barland2022displacement}. Hilbert transformation (as in \cite{arriaga2014speckle}) would require careful windowing for real time operation, impacting operational bandwidth. Alternatives include original phase unwrapping approaches \cite{orakzai2022fast,ri2021vibration} but also in this case experimental observations are limited to spectrally narrow displacements, while neural networks operate over a decade-broad spectrum (limited by labelled training data, see also \cite{barland2021convolutional}). 
One key reference indicating real time inference is \cite{magnani2012self} but experimental validation on non-periodic displacement is scarce. Thus, without removing any merit to existing alternatives, key assets of direct displacement inference by neural networks include adaptability to spectrally broad displacements through convolutions and depth without critical parameter tuning or estimation \cite{barland2021convolutional}, robustness to noise and adaptability to poor signal quality \cite{barland2021convolutional,matha2023high} and now real time inference at kHz rate with a few tens of mW computing power, opening the way to robust autonomous sensors. 

\section{Conclusion}
\label{sec:conclusion}

In this article, we presented a first prototype of an embedded device for small displacement measurement through self-mixing interferometry.
To achieve this, a CNN and a ResNet deep neural networks compatible with the embedded constraints were designed and evaluated.
The neural networks have been made robust against variations of amplitude of the input signal by adding a per-window normalization layer.

Then, the neural networks were quantized in order to reduce the memory footprint and inference latency.
Finally, the neural networks were deployed on an STM32L476RG and an STM32U575ZI microcontrollers.
The memory footprint and latency were evaluated on both platforms.

Results showed that the signal reconstruction with the ResNet could achieve a 0.8893 correlation with the ground truth (mean absolute error of 0.35$\lambda/ms$) and that the memory footprint is kept below 10\% of the microcontroller capabilities.
When quantized onto 16 bits and deployed on the STM32U575ZI microcontroller, the latency stays below the real-time constraint of one millisecond, therefore enabling real-time reconstruction. With this real time operation rate, a displacement with frequency up to 500Hz could in principle be measured (from the sampling theorem). However, 
It is worth reminding that the validity of the Neural Network approach we use here is bound by the domain covered by the training data outlined in \cite{barland2021convolutional}: weak to moderate feedback (no multistability in the response, C near unity), target speed below 2.5~$\lambda/ms$ and target motion frequency between 5 and 100Hz. The accuracy would also be ultimately bound by the ground truth accuracy of the training set but in this case the main limitation comes from the neural network inference itself.

While the current platform can perform the processing with a 48~kHz sampling rate, higher sampling rates would provide more accurate reconstruction while also improving the frequency range.
For the neural network to process the data at a fast enough pace, a hardware accelerator could be used instead of a microcontroller.

Future works will also focus on optimizing the energy consumption further, especially the low-noise amplifier, in order to provide an autonomous sensor.

Finally, the embedded platform currently does not provide any quality of service information. Training a dedicated neural network providing both displacement inference and quality of service estimation is currently under investigation.

Beyond that, this work opens the way to many applications of AI-enabled self-mixing sensors such as flow cytometry, microfluidics or environmental conditions classification.

\bibliographystyle{IEEEtran}
\bibliography{bibliography}

\end{document}